\documentclass[a4paper,11pt]{article}
\usepackage{pos}

\title{GPDs in asymmetric frames}

\author*[a]{Shohini Bhattacharya}
\author[b]{Krzysztof Cichy}
\author[c]{Martha Constantinou}
\author[c]{Jack Dodson}
\author[d]{Xiang Gao}
\author[c]{Andreas Metz}
\author[a]{Swagato Mukherjee}
\author[c]{Aurora Scapellato}
\author[e]{Fernanda Steffens}
\author[d]{Yong Zhao}

\affiliation[a]{Brookhaven National Laboratory, \\
Upton, New York 11973, USA}

\affiliation[b]{Adam Mickiewicz University, ul.\ Uniwersytetu Pozna\'nskiego 2 \\
61-614 Pozna\'{n}, Poland}

\affiliation[c]{Temple University, \\
Philadelphia,  PA 19122 - 1801,  USA}

\affiliation[d]{Argonne National Laboratory, \\
Lemont, IL 60439, USA}

\affiliation[e]{Institut f\"ur Strahlen- und Kernphysik, Rheinische Friedrich-Wilhelms-Universit\"at Bonn \\
Nussallee 14-16, 53115 Bonn}

\emailAdd{sbhattach@bnl.gov}

\abstract{It is often taken for granted that Generalized Parton Distributions (GPDs) are defined in the "symmetric" frame, where the transferred momentum is symmetrically distributed between the incoming/outgoing hadrons. However, such frames pose computational challenges for the lattice QCD practitioners. In these proceedings, we lay the foundation for lattice QCD calculations of GPDs in "asymmetric" frames, where the transferred momentum is not symmetrically distributed between the incoming/outgoing hadrons. The novelty of our work relies on the parameterization of the matrix elements in terms of Lorentz-invariant amplitudes, which not only helps in establishing relations between the said frames but also helps in isolating higher-twist contaminations. As an example, we focus on the unpolarized GPDs for spin-1/2 particles.}

\FullConference{%
  The 39th International Symposium on Lattice Field Theory (Lattice2022),\\
  8-13 August, 2022 \\
  Bonn, Germany 
}

\begin{document}
\maketitle

\section{Introduction}
Generalized Parton Distributions (GPDs) are the 3D generalizations of the collinear Parton Distribution Functions (PDFs)~\cite{Ji:1996ek,Radyushkin:1996nd}. There are several motivations to study GPDs:
\begin{itemize}
    \item For $\xi=0$ the Fourier transforms of the GPDs are related to the impact-parameter distributions which provide information about the three-dimensional distribution of partons --- (one-dimensional) longitudinal momentum distribution; (two-dimensional) transverse spatial distribution, see for example Ref.~\cite{Burkardt:2000za}.
    
    \item Twist-2 GPDs are related to the total angular momentum of partons~\cite{Ji:1996ek}.
    
    \item One should look for other ways to access GPDs because of the challenges involved in their extraction through the processes of Deep Virtual Compton Scattering (DVCS)~\cite{Radyushkin:1996nd} and meson production~\cite{Collins:1996fb}. Challenges are caused by the sensitivity of differential cross-sections to only $x$-integrals of GPDs, and not GPDs themselves~\cite{Ji:1996ek, Radyushkin:1996nd}. Therefore, it is desirable to extract the $x$-dependence of the GPDs from first principles within Lattice QCD. However, for a very long time this was not possible because of time-dependence of these quantities. As a result, all of the lattice calculations were limited to the calculations of lowest Mellin moments of the GPDs, see Ref.~\cite{Constantinou:2014tga}. In 2013, there was a path-breaking proposal by X. Ji to calculate instead auxiliary quantities called "quasi-GPDs"~\cite{Ji:2013dva,Ji:2014gla,Ji:2020ect}. This approach relies on the extraction of matrix elements for boosted hadrons involving spatially-separated fields. Ever since this proposal, enormous progress has taken place, see some reviews~\cite{Constantinou:2020pek,Cichy:2021lih,Cichy:2021ewm}. In fact, Ref.~\cite{Alexandrou:2020zbe} provides the first-ever lattice-QCD results of the unpolarized and helicity GPDs of the nucleon from the quasi-distribution approach. Lattice QCD calculations have the potential to not only provide insight into the experimentally-inaccessible features of GPDs, but also help in extracting the "full" GPDs from the existing experimental data. 
\end{itemize}

\section{Formalisms to calculate GPDs in asymmetric frames}
\label{sec:strategy}
\subsection{Frames: Symmetric and asymmetric}
\label{s:frames}
\begin{figure}[t]
\centering
\fbox{\includegraphics[width = 5cm]{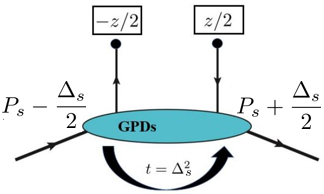}}
\hspace{1cm}
\fbox{\includegraphics[width = 5cm]{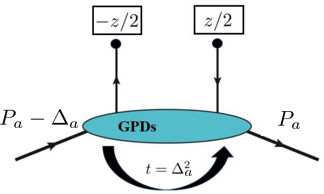}}
\caption{Graphical representation of the two frames employed in this work.
Left plot: Symmetric frame. Right plot: Asymmetric frame.}
\label{fig}
\end{figure}
The most widely used frame of reference to calculate GPDs is the symmetric frame. For this frame, the momentum transfer is symmetrically distributed between the incoming ($p_i$) and the outgoing hadrons ($p_f$) (see left plot of Fig.~\ref{fig}). However, one can also think of a frame where the momentum transfer is not equally shared between the incoming and outgoing hadrons, but is rather exclusively applied to the incoming hadron (see, right plot of Fig.~\ref{fig}). Such a frame is known as an asymmetric frame.

Lattice calculations of GPDs has primarily been confined to symmetric frames. However, such frames pose serious computational challenges because they require separate calculation for each values of the momentum transfer ($\Delta$), resulting in increased computational costs. So the question that we strive to address in this work is: Can we lay a formalism to systematically perform lattice calculations of GPDs in asymmetric frames (which is expected to be computationally less expensive)? In this work, we argue that there are two approaches to solving this question. In the first approach, we will show that it is possible to relate the two frames via an appropriate Lorentz transformation. In the second approach, we will propose a Lorentz-covariant decomposition of the lattice matrix elements in terms of Lorentz-invariant (frame-independent) amplitudes. These amplitudes will then be used to make connections between the two frames. As a byproduct, we will show that this approach helps in identifying higher-twist contaminations which may be present in quasi-GPDs at finite values of momentum.

\subsection{Lorentz transformation approach}
\label{s:LT}
In this section, we explain the Lorentz transformation (LT) approach. First, it is straight-forward to realize that a LT along the $z$-direction is not optimal for lattice calculations because this requires a spatial operator distance (say $z = (0,0_\perp,z^3 \neq 0)$) to pick up a temporal component (that is $z \xrightarrow[]{\text{LT}} (z^0 \neq 0, 0_\perp, z^3)$).
However, a LT applied to any direction transverse to the $z$-axis does not change the spatial nature of operator distances. This transformation is called as the "transverse boost". We explain this by considering a transverse boost in the $x$-direction and for the simplest case of zero skewness. The logic can be generalized for any general transverse boost and for arbitrary values of skewness.

We begin by relating the incoming state in the two frames, $p_i^s = (E_i^s, - \Delta^{1,s}/2,0,P^3)$ and $p_i^a=(E_i^a, -\Delta^{1,a}, 0, P^3)$. 
LT provides $p^s = \Lambda_{\rm LT} \, p^a$,
\begin{align}
\begin{pmatrix}
E_i^s \\[0.1cm]
p_i^{1,s} \\[0.1cm]
p_i^{2,s} \\[0.1cm]
p_i^{3,s}
\end{pmatrix}
& =
\begin{pmatrix}
\gamma & -\gamma \beta & 0 & 0 \\[0.1cm]
-\gamma \beta & \gamma & 0 & 0 \\[0.1cm]
0 & 0 & 1 & 0 \\[0.1cm]
0 & 0 & 0 & 1
\end{pmatrix}
\times 
\begin{pmatrix}
E_i^a \\[0.1cm]
-\Delta^{1,a} \\[0.1cm]
0 \\[0.1cm]
P^3
\end{pmatrix} \, .
\label{e:LT}
\end{align}
This gives,
\begin{align}
    E_i^s & = \gamma (E_i^a+\beta \Delta^{1,a}) \, ,
\label{e:LT1}
\end{align}
and,
\begin{align}
 p_i^{1,s} & = - \gamma (\beta E_i^a +\Delta^{1,a}) \quad \rightarrow \quad  \Delta^{1,s} = 2\gamma (\beta E_i^a +\Delta^{1,a}) \, .
 \label{e:LT2}
\end{align}
Similarly, the outgoing state in the two frames, $p_f^s = (E_f^s, \Delta^{1,s}/2,0,P^3)$ and $p_f^a=(E_f^a, 0, 0, P^3)$ can also be related. (Keep in mind that the energies of the incoming and outgoing states are different in the asymmetric frame.) We then find,
\begin{align}
E_i^s & = \gamma E_f^a \, ,
\label{e:LT3}
\end{align}
and,
\begin{align}
 p_f^{1,s} & = -\gamma \beta E_f^a \quad \rightarrow \quad 
 \Delta^{1,s} = -2\gamma \beta E_f^a \, .
 \label{e:LT4}
\end{align}
From Eqs.~(\ref{e:LT1}) and~(\ref{e:LT3}), we find,
\begin{align}
\beta & = - \bigg ( \dfrac{E_i^a-E_f^a}{\Delta^{1,a}} \bigg ) \, .
\label{e:b1}
\end{align}
From Eqs.~(\ref{e:LT2}) and~(\ref{e:LT4}), we find,
\begin{align}
\beta & = - \dfrac{\Delta^{1,a}}{E_i^a+E_f^a} \, .
\label{e:b2}
\end{align}
Then, Eqs.~(\ref{e:b1}) and~(\ref{e:b2}) imply,
\begin{align}
\Delta^{1,a} = \sqrt{(E_i^a)^2-(E_f^a)^2} \, .
\end{align}
Hence, $\beta$ can be written as,
\begin{align}
\beta = - \sqrt{\dfrac{E_i^a-E_f^a}{E_i^a+E_f^a}} <0 \, .
\label{e:beta}
\end{align}
This implies $\Delta^{0,a} < 0$, and 
\begin{align}
\gamma = \dfrac{1}{\sqrt{1-\beta^2}} = \sqrt{\dfrac{E_i^a+E_f^a}{2E_f^a}} \, .
\label{e:gamma}
\end{align}
Therefore, by using the expressions for $(\beta, \gamma)$, we can write down uniquely the symmetric frame variables $(E_i^s, \Delta^{1,s})$ in terms of the asymmetric frame variables $(E_i^a, E_f^a, \Delta^{1,a})$: The energy should be,
\begin{align}
E_i^s = \gamma E_f^a = \sqrt{\dfrac{E_f^a(E_i^a+E_f^a)}{2}} \, ,
\end{align}
and the transverse-momentum transfer,
\begin{align}
\Delta^{1,s} & = - 2\gamma \beta E_f^a \, ,\nonumber \\[0.2cm]
\text{or,} \quad \Delta^{1,s} & = 2\sqrt{\dfrac{E_f^a(E_i^a-E_f^a)}{2}} = 2 \sqrt{\dfrac{E_f^a}{2(E_i^a+E_f^a)}} \,  \Delta^{1,a} \, .
\end{align}
We repeat that the above method can be generalized for $\vec{\Delta}_{\perp} = (\Delta^1, \Delta^2)$ and for arbitrary values of skewness. 

Now that we have sketched the idea of how to relate the kinematical variables between the two frames, we proceed to understand how the matrix elements defining quasi-GPDs transform between the two frames. For this purpose, we focus on spin-0 particles such as the pion. (The method can be generalized for spin-1/2 particles.) The (unpolarized) pion GPD is defined as,
\begin{align}
F^{\mu}(z, P, \Delta) & = \langle p_f | \bar{q} (-\tfrac{z}{2}) \gamma^\mu \, {\cal W}(-\tfrac{z}{2}, \tfrac{z}{2})  q (\tfrac{z}{2}) | p_i\rangle \, .
\label{e:mat}
\end{align}
Here, ${\cal W}$ is a straight Wilson line required to make the correlator gauge invariant. Historically, (unpolarized) quasi-GPDs have been defined through matrix elements of the operator $\gamma^0$, see for instance Refs.~\cite{ Radyushkin:2019owq,Alexandrou:2020zbe}. By applying the transverse boost Eq.~(\ref{e:LT}), we find that the matrix element $\langle .. \gamma^0 .. \rangle $ in the symmetric frame can be expressed in terms of matrix elements of different operators $\langle .. (\gamma^0 + \gamma^1 ) .. \rangle $ in the asymmetric frame,
\begin{align}
\langle p_f | \bar{q} (-\tfrac{z}{2}) \gamma^0 \, {\cal W}(-\tfrac{z^3}{2}, \tfrac{z^3}{2}) \, q (\tfrac{z}{2}) |p_i \rangle^s & = \gamma \langle p_f | \bar{q} (-\tfrac{z}{2}) \gamma^0 \, {\cal W}(-\tfrac{z^3}{2}, \tfrac{z^3}{2}) \,  q (\tfrac{z}{2}) |p_i \rangle^a \nonumber \\[0.2cm]
& - \gamma \beta \langle p_f | \bar{q} (-\tfrac{z}{2}) \gamma^1 \, {\cal W}(-\tfrac{z^3}{2}, \tfrac{z^3}{2}) \, q (\tfrac{z}{2}) |p_i \rangle^a .
\label{e:LT_matrix_elements}
\end{align}
This equation simply reflects how the $0^{\rm th}$ component of a 4-vector changes under the Lorentz transformation Eq.~(\ref{e:LT}). Therefore, this implies that a transverse boost that fixes $(\beta,\gamma)$ (Eqs.~(\ref{e:beta}) and~(\ref{e:gamma})) allows for an exact calculation of quasi-GPDs in the symmetric frame through matrix elements of the asymmetric frame. However, Eq.~(\ref{e:LT_matrix_elements}) also shows that a quasi-GPD defined through the operator $\gamma^0$ is not Lorentz invariant. In the limit of a large momentum, we recover,
\begin{align}
\lim_{P^{3} \rightarrow \infty} \langle .. \gamma^0 .. \rangle^s &\, \approx\, \langle .. \gamma^0 .. \rangle^a + \mathcal{O} \bigg ( \dfrac{1}{P^{3}} \bigg ) \langle .. \gamma^1 .. \rangle^a  \,\rightarrow\, \langle .. \gamma^0 .. \rangle^a \, ,
\end{align}
which means that the contribution from the matrix element $\langle .. \gamma^1 .. \rangle $ maybe viewed as a power correction at finite values of momentum $P^3$.

\subsection{Amplitude approach: Spin-$\boldsymbol{1/2}$ particles}
\label{s:amp}
In this section, we explain the amplitude approach through the example of spin-1/2 particles, such as the proton. (We refer to Ref.~\cite{Bhattacharya:2022aob} for details on spin-0 particles.) As a first step, we build a Lorentz-covariant decomposition of the vector matrix element in terms of the available vectors $(P^{\mu}, z^{\mu}, \Delta^\mu)$.
By considering constraints from parity, we find that the general structure of the vector matrix element involves eight linearly-independent Dirac structures multiplied by eight Lorentz-invariant (frame-independent) amplitudes, 
\begin{align}
\label{eq:parametrization_general}
F^{\mu} (z,P,\Delta) & = \bar{u}(p_f,\lambda') \bigg [ \dfrac{P^{\mu}}{m} A_1 + m z^{\mu} A_2 + \dfrac{\Delta^{\mu}}{m} A_3 + i m \sigma^{\mu z} A_4 + \dfrac{i\sigma^{\mu \Delta}}{m} A_5 \nonumber \\
& + \dfrac{P^{\mu} i\sigma^{z \Delta}}{m} A_6 + m z^{\mu} i\sigma^{z \Delta} A_7 + \dfrac{\Delta^{\mu} i\sigma^{z \Delta}}{m} A_8  \bigg ] u(p_i, \lambda) \, .
\end{align}
Here
$\sigma^{\mu \nu} \equiv \tfrac{i}{2} (\gamma^\mu \gamma^\nu - \gamma^\nu \gamma^\mu)$,  
$\sigma^{\mu z} \equiv \sigma^{\mu \rho} z_\rho$, 
$\sigma^{\mu \Delta} \equiv \sigma^{\mu \rho} \Delta_\rho$, $\sigma^{z \Delta} \equiv \sigma^{\rho \tau} z_\rho \Delta_\tau$, 
$z \equiv (z^0 = 0, z_\perp = 0_\perp, z^3 \neq 0)$. 
(For a derivation of Eq.~(\ref{eq:parametrization_general}), we refer to Ref.~\cite{Meissner:2009ww}. See also Ref.~\cite{Rajan:2017cpx} where the vector matrix element has been parameterized in the momentum space for a straight Wilson line.) 
For brevity, we use the compact notation $A_i \equiv A_i (z\cdot P, z \cdot \Delta, \Delta^2, z^2)$, with $A_i$'s being the Lorentz-invariant amplitudes whose arguments are functions of Lorentz scalars\footnote{In the literature, the amplitudes have also been called generalized Ioffe time distributions (ITDs)~\cite{Radyushkin:2019owq}.}.

For spin-$1/2$ particles, the vector matrix element can be parameterized in terms of two light-cone GPDs $H$ and $E$~\cite{Diehl:2002he},
\begin{align}
F^{+} (z, P^{s/a}, \Delta^{s/a}) & = \bar{u}^{s/a}(p_f^{s/a}, \lambda ') \bigg [\gamma^{+} H(z, P^{s/a}, \Delta^{s/a})  \nonumber \\
& \hspace{3cm} + \frac{i\sigma^{+\mu}\Delta^{s/a}_{\mu}}{2m} E(z, P^{s/a}, \Delta^{s/a}) \bigg ] u^{s/a}(p^{s/a}_i, \lambda) \, .
\label{e:GPD_para_spin1/2}
\end{align}
By using $\mu =+$ in Eq.~(\ref{eq:parametrization_general}), followed by a subsequent change of basis, it is possible to map the $A_i$'s onto the $H$ and $E$ GPDs in Eq.~(\ref{e:GPD_para_spin1/2}). The results are,
\begin{align}
\label{eq:H}
H (z,P^{s/a},\Delta^{s/a}) & = A_1 + \dfrac{\Delta^{+,s/a}}{P^{+,s/a}} A_3 \, , \\
\label{eq:E}
E (z,P^{s/a},\Delta^{s/a}) & = - A_1 - \dfrac{\Delta^{+,s/a}}{P^{+,s/a}} A_3 + 2 A_5 + 2P^{+,s/a}z^- A_6 + 2 \Delta^{+,s/a} z^- A_8 \, .
\end{align}
Keep in mind that the arguments of the $A_i$'s for light-cone GPDs have no dependence on $z^{2}$. Also, $z^\mu=(0,z^-,0_\perp)$ and $\Delta^+/P^+=z\cdot \Delta / z\cdot P$, etc. Thus, it is possible to write the above expressions in a Lorentz invariant way as,
\begin{align}
\label{eq:H_improved}
H  (z\cdot P^{s/a}, z \cdot \Delta^{s/a}, (\Delta^{s/a})^2)& = A_1 + \dfrac{\Delta^{s/a} \cdot z}{P^{s/a} \cdot z} A_3 \, , \\
\label{eq:E_improved}
E(z\cdot P^{s/a}, z \cdot \Delta^{s/a}, (\Delta^{s/a})^2) & = - A_1 - \dfrac{\Delta^{s/a} \cdot z}{P^{s/a} \cdot z} A_3 + 2 A_5 + 2 P^{s/a} \cdot z A_6 + 2\Delta^{s/a} \cdot z A_8 \, .
\end{align}
This means the light-cone GPDs are frame-independent as long as the Lorentz scalars $(z \cdot P^{s/a}, z \cdot \Delta^{s/a}, (\Delta^{s/a})^2)$ are the same in the two frames.

Next, we turn to the quasi-GPDs $\mathcal{H}$ and $\mathcal{E}$, which historically have been defined in terms of matrix elements of $\gamma^0$ operator as~\cite{Bhattacharya:2018zxi,Bhattacharya:2019cme},
\begin{align}
F^{0} (z, P^{s/a},\Delta^{s/a}) & = \langle p_f^{s/a}, \lambda' | \bar{q} (-\tfrac{z}{2}) \gamma^0 q (\tfrac{z}{2}) | p^{s/a}_i, \lambda\rangle \nonumber \\[0.1cm]
& = \bar{u}^{s/a}(p_f^{s/a}, \lambda ') \bigg [\gamma^{0} {\cal{H}}^{s/a}_0 (z,P^{s/a},\Delta^{s/a}) \nonumber \\
& \hspace{3cm}  + \frac{i\sigma^{0\mu}\Delta^{s/a}_{\mu}}{2m} {\cal{E}}^{s/a}_0 (z,P^{s/a},\Delta^{s/a}) \bigg ] u^{s/a}(p^{s/a}_i, \lambda) \, .
\label{e:historic}
\end{align}
If we use $\mu =0$ in Eq.~(\ref{eq:parametrization_general}), then after performing a change of basis it is possible to map the $A_i$'s onto the quasi-GPDs in Eq.~(\ref{e:historic}). The relations in the symmetric frame read,
\begin{align}
\label{eq:quasiH_symm}
{\cal{H}}^{s}_0 (z,P^{s},\Delta^{s}) & = A_1 + \dfrac{\Delta^{0,s}}{P^{0,s}} A_3 - \dfrac{m^{2} \Delta^{0,s} z^3}{2P^{0,s} P^{3,s}} A_4 + \bigg [ \dfrac{(\Delta^{0,s})^{2} z^{3}}{2P^{3,s}}  - \dfrac{\Delta^{0,s} \Delta^{3,s} z^3 P^{0,s}}{2(P^{3,s})^2} - \dfrac{z^{3} (\Delta^{s}_\perp)^2}{2P^{3,s}} \bigg ] A_6 \nonumber \\
& +  \bigg [ \dfrac{(\Delta^{0,s})^{3} z^{3}}{2P^{0,s} P^{3,s}}  - \dfrac{(\Delta^{0,s})^2 \Delta^{3,s} z^3}{2(P^{3,s})^2} - \dfrac{\Delta^{0,s} z^{3} (\Delta^{s}_\perp)^2}{2P^{0,s} P^{3,s}} \bigg ] A_8 \, ,
\\[0.2cm]
\label{eq:quasiE_symm}
{\cal{E}}^{s}_0 (z,P^{s},\Delta^{s}) & = - A_1 - \dfrac{\Delta^{0,s}}{P^{0,s}} A_3 + \dfrac{m^2 \Delta^{0,s} z^{3}}{2P^{0,s} P^{3,s}} A_4 + 2 A_5 + \bigg [ - \dfrac{(\Delta^{0,s})^{2}z^{3}}{2P^{3,s}}  + \dfrac{P^{0,s} \Delta^{0,s} \Delta^{3,s} z^{3}}{2 (P^{3,s})^{2}} + \dfrac{z^{3} (\Delta^{s}_\perp)^2}{2 P^{3,s}} \nonumber \\
& - \dfrac{2z^{3}(P^{0,s})^{2}}{P^{3,s}} \bigg ] A_6 + \bigg [ - \dfrac{(\Delta^{0,s})^{3}z^{3}}{2P^{0,s} P^{3,s}}  + \dfrac{(\Delta^{0,s})^2 \Delta^{3,s} z^{3}}{2(P^{3,s})^{2}} + \dfrac{\Delta^{0,s} z^{3}(\Delta^{s}_\perp)^2}{2P^{0,s} P^{3,s}} - \dfrac{ 2z^{3}P^{0,s} \Delta^{0,s}}{P^{3,s}} \bigg ] A_8 \, .
\end{align}
On the other hand, the relations in the asymmetric frame read, 
\begin{align}
\label{eq:quasiH_nonsymm}
& {\cal{H}}^{a}_0 (z,P^{a},\Delta^{a}) = A_1 + \dfrac{\Delta^{0,a}}{{P}^{0,a}} A_3 - \bigg [ \dfrac{m^2 \Delta^{0,a} z^3}{2{P}^{0,a} {P}^{3,a}} - \dfrac{1}{(1+\tfrac{\Delta^{3,a}}{2{P}^{3,a}})} \dfrac{m^2 \Delta^{0,a} \Delta^{3,a} z^3}{4 {P}^{0,a} ({P}^{3,a})^2} \bigg ] A_4 \nonumber \\[1ex]
& + \bigg [ \dfrac{(\Delta^{0,a})^2 z^3}{2{P}^{3,a}} - \dfrac{1}{(1+\tfrac{\Delta^{3,a}}{2{P}^{3,a}})} \dfrac{(\Delta^{0,a})^2 \Delta^{3,a} z^3}{4 ({P}^{3,a})^2} - \dfrac{1}{(1+\tfrac{\Delta^{3,a}}{2{P}^{3,a}})} \dfrac{{P}^{0,a} \Delta^{0,a} \Delta^{3,a} z^3}{2 ({P}^{3,a})^2} - \dfrac{z^3 (\Delta^{a}_\perp)^2}{2 {P}^{3,a}} \bigg ] A_6
\nonumber\\[1ex]
& + \bigg [ \dfrac{(\Delta^{0,a})^3 z^3}{2{P}^{0,a} {P}^{3,a}} - \dfrac{1}{(1+\tfrac{\Delta^{3,a}}{2{P}^{3,a}})} \dfrac{(\Delta^{0,a})^3 \Delta^{3,a} z^3}{4 {P}^{0,a} ({P}^{3,a})^2} - \dfrac{1}{(1+\tfrac{\Delta^{3,a}}{2{P}^{3,a}})} \dfrac{(\Delta^{0,a})^2 \Delta^{3,a} z^3}{2({P}^{3,a})^2} - \dfrac{z^3 (\Delta^{a}_\perp)^2 \Delta^{0,a}}{2 {P}^{0,a} {P}^{3,a}} \bigg ] A_8 \, ,
\\[3ex]
\label{eq:quasiE_nonsymm}
& {\cal{E}}^{a}_0 (z,P^{a},\Delta^{a}) = - A_1 - \dfrac{\Delta^{0,a}}{{P}^{0,a}} A_3 -  \bigg [ - \dfrac{m^2 \Delta^{0,a} z^3}{2 {P}^{0,a} {P}^{3,a}} - \dfrac{1}{(1 + \tfrac{\Delta^{3,a}}{2 {P}^{3,a}})} \bigg ( \dfrac{m^2 z^3}{{P}^{3,a}} - \dfrac{m^2 \Delta^{0,a} \Delta^{3,a} z^3}{4 {P}^{0,a} ({P}^{3,a})^2} \bigg ) \bigg ] A_4 + 2A_5 \nonumber \\[1ex]
& + \bigg [ - \dfrac{(\Delta^{0,a})^2 z^{3}}{2 {P}^{3,a}} - \dfrac{1}{(1 +\tfrac{\Delta^{3,a}}{2{P}^{3,a}})} \bigg ( \dfrac{{P}^{0,a} \Delta^{0,a} z^3}{ {P}^{3,a}} - \dfrac{(\Delta^{0,a})^2 \Delta^{3,a} z^3}{4 ({P}^{3,a})^2} \bigg ) - \dfrac{1}{(1+\tfrac{\Delta^{3,a}}{2 {P}^{3,a}})} \bigg ( \dfrac{2z^3 ({P}^{0,a})^2}{{P}^{3,a}} 
\nonumber\\[1ex]
& - \dfrac{{P}^{0,a} \Delta^{0,a} \Delta^{3,a} z^3}{2 ({P}^{3,a})^2} \bigg ) + \dfrac{z^3 (\Delta^{a}_\perp)^2}{2{P}^{3,a}} \bigg ] A_6 + \bigg [ - \dfrac{(\Delta^{0,a})^3 z^{3}}{2 {P}^{0,a}{P}^{3,a}} - \dfrac{1}{(1 +\tfrac{\Delta^{3,a}}{2{P}^{3,a}})} \bigg ( \dfrac{ (\Delta^{0,a})^2 z^3}{{P}^{3,a}} - \dfrac{(\Delta^{0,a})^3 \Delta^{3,a} z^3}{4 \overline{P}^{0,a} ({P}^{3,a})^2} \bigg ) \nonumber \\[1ex]
& - \dfrac{1}{(1+\tfrac{\Delta^{3,a}}{2 {P}^{3,a}})} \bigg ( \dfrac{2z^3 {P}^{0,a} \Delta^{0,a}}{{P}^{3,a}} - \dfrac{(\Delta^{0,a})^2 \Delta^{3,a} z^3}{2 ({P}^{3,a})^2} \bigg ) + \dfrac{z^3 (\Delta^{a}_\perp)^2 \Delta^{0,a}}{2 {P}^{0,a} {P}^{3,a}} \bigg ] A_8 \, .
\end{align}

However, one can think of other definitions of quasi-GPDs. For this purpose, we recall the position-space matching relation between, for instance, light-cone GPD $H$ and quasi-GPD $\mathcal{H}$~\cite{Radyushkin:2019owq}:
\begin{align}\label{eq:match}
{\cal{H}} \big(z \cdot P, -2\xi(z\cdot P), \Delta^2, z^{2}, \mu^2 \big) & = \int_{-1}^1 du\, \bar{C}\, (u, z \cdot P, \xi, z^2, \mu^2) \, H \big(u(z \cdot P), -2u\xi(z\cdot P), \Delta^2, \mu^2\big) \, .
\end{align}
Here, $\bar{C}$ is the pertubatively-calculable matching coefficient~\cite{Radyushkin:2019owq} and $\mu$ is the renormalization scale in the $\overline{\rm MS}$ scheme. 
At leading order in $\alpha_s$, the above formula indicates that ${\cal{H}}$ collapses to $H$ in the light-cone limit $z^2\to0$,
\begin{align}
\lim_{z^2\to 0} {\cal{H}} (z\cdot P, z \cdot \Delta, \Delta^2, z^2) = H (z\cdot P, z \cdot \Delta, \Delta^2,0) + {\cal O}(\alpha_s) \, .
\label{e:spin0_LO_arg}
\end{align}
Therefore, a natural way to define the quasi-GPDs ${\cal{H}}$ and ${\cal{E}}$ is through a Lorentz-invariant generalization of the light-cone definitions in Eqs.~(\ref{eq:H_improved}) and~(\ref{eq:E_improved}) to $z^{2} \neq 0$, i.e.,
\begin{align}
\label{eq:Hq_improved}
{\cal H}  (z\cdot P^{s/a}, z \cdot \Delta^{s/a}, (\Delta^{s/a})^2, z^{2})& = A_1 + \dfrac{\Delta^{s/a} \cdot z}{P^{s/a} \cdot z} A_3 \, , \\[0.2cm]
\label{eq:Eq_improved}
{\cal E}(z\cdot P^{s/a}, z \cdot \Delta^{s/a}, (\Delta^{s/a})^2 ,z^2) & = - A_1 - \dfrac{\Delta^{s/a} \cdot z}{P^{s/a} \cdot z} A_3 + 2 A_5 + 2 P^{s/a} \cdot z A_6 + 2 \Delta^{s/a} \cdot z  A_8 \, ,
\end{align}
where now the arguments of the $A_i$'s have a non-zero dependence on $z^{2}$. We expect the definitions in
Eqs.~(\ref{eq:Hq_improved}-(\ref{eq:Eq_improved}) to have two advantages: First, these definitions may converge faster to the light-cone
GPDs because of the similarities in their functional forms with their (respective) light-cone GPDs.
(Such a statement is inspired from Ref.~\cite{Radyushkin:2017cyf}, where similar arguments were made for the quasi-PDFs.
See also the next paragraph for explicit explanations.) Second, these definitions differ from their
light-cone GPDs by frame-independent power corrections; contrast with historic definitions which are frame-dependent.

We now discuss in detail the various definitions of quasi-GPDs: We notice that for finite values of the momentum, the historic definitions of quasi-GPDs (${\cal{H}}^{s/a}_0 (A_i; z) \, , {\cal{E}}^{s/a}_0 (A_i; z)$) in Eqs.~(\ref{eq:quasiH_symm})-(\ref{eq:quasiE_nonsymm}) involve additional amplitudes that are not present in the light-cone GPDs, Eqs.~(\ref{eq:H_improved})-(\ref{eq:E_improved}). This is not the case for the Lorentz-invariant definitions of quasi-GPDs (${\cal{H}} (A_i; z) \, , {\cal{E}} (A_i; z)$) in Eqs.~(\ref{eq:Hq_improved})-(\ref{eq:Eq_improved}). (Note that this is different from the (unpolarized) quasi-PDF case where arguments were made in favor of $\gamma^0$ (against $\gamma^3$) because of the absence of such additional amplitudes relative to the (unpolarized) light-cone PDF case~\cite{Radyushkin:2017cyf}.) Therefore, the additional amplitudes in (${\cal{H}}^{s/a}_0 (A_i; z) \, , {\cal{E}}^{s/a}_0 (A_i; z)$) may be viewed as contaminations from explicit power corrections, which one would have to suppress by going to larger and larger values of momentum. Hence, we believe that (${\cal{H}} (A_i; z) \, , {\cal{E}} (A_i; z)$) may converge relatively faster to their (respective) light-cone GPDs, simply because of the absence of such additional amplitudes. (Of course, (${\cal{H}} (A_i; z) \, , {\cal{E}} (A_i; z)$) also have power corrections, but they are implicit within the amplitudes themselves. Our argument above is for the power corrections that are explicit.) Our reasoning is perhaps too simple and for sure needs further substantiation. In fact, it may be that the actual convergence of the various definitions of quasi-GPDs is determined by the underlying dynamics. Note that the Lorentz non-invariance of the historical definitions of quasi-GPDs implies that the basis vectors $(\gamma^0, i\sigma^{0\Delta^{s/a}})$ do not form a complete set for spatially-separated bi-local operators for finite values of momentum. Therefore, we can argue that the Lorentz-invariant definitions are in fact just a redefinition of quasi-GPDs in terms of a suitable linear combination of operators (which turns out to be $\gamma_\perp$) that make them functions of Lorentz scalars~\cite{Bhattacharya:2022aob}.

In Ref.~\cite{Bhattacharya:2022aob} and~\cite{Constantinou:2022fqt}, we compare numerically the different definitions of quasi-GPDs for $\xi =0$ to get an idea about the relative size of power corrections. Finally, we remark on the matching coefficient for the different definitions of quasi-GPDs: It is known that the GPD matching coefficient for the operator $\gamma^0$ reduces to that for the corresponding PDF when $\xi =0$, even if $t \neq 0$~\cite{Radyushkin:2019owq}. The PDF matching coefficient for $\gamma^0$ is for the amplitude $A_1$, which is also the only contributing amplitude to the LI definition of the GPD when $\xi =0$. Therefore, the matching coefficients for the $\gamma^0$ and the LI definitions of the GPDs are equal. We will elaborate this point more, including the general case of $\xi \neq 0$, in a forthcoming publication.

\section{Summary}
In these proceedings, we have laid down the theoretical tools to perform lattice QCD calculations of GPDs in asymmetric frames. We have highlighted two approaches to performing such calculations:
\begin{itemize}
    \item Lorentz transformation (LT) approach (Sec.~\ref{s:LT}): We have shown that there exists a LT called the "transverse boost" (transverse with respect to the Wilson Line) that allows one to uniquely relate the kinematical variables as well as the matrix elements in the two frames. 
    \item Amplitude approach (Sec.~\ref{s:amp}): We have proposed a Lorentz-covariant decomposition of the vector matrix element in terms of Lorentz-invariant/frame-independent amplitudes. The amplitudes can be used as tools to relate the two frames. This approach also shows that at finite values of the boost momentum the historic definitions of quasi-GPDs (defined through $\gamma^0$) have additional amplitudes that are not present in the light-cone limit. This motivates us to come up with alternative definitions of quasi-GPDs that may potentially converge faster. One such candidate can be the case where one chooses the same functional form as the light-cone GPDs subjected to include $z^2 \neq 0$. Naively, because of the similarity in the functional forms (or because of the absence of additional amplitudes), one may expect such a definition of quasi-GPD to converge faster to the light-cone GPD. Such a definition is also frame-independent, contrary to the historic definitions.
\end{itemize}

\begin{center}
    \textbf{Acknowledgements}
\end{center}
This material is based upon work supported by the U.S. Department of Energy, Office of Science, Office of Nuclear Physics through Contract No.~DE-SC0012704, No.~DE-AC02-06CH11357 and within the framework of Scientific Discovery through Advance Computing (SciDAC) award Fundamental Nuclear Physics at the Exascale and Beyond (S.~B. and S.~M.).
K.~C.\ is supported by the National Science Centre (Poland) grants SONATA BIS no.\ 2016/22/E/ST2/00013 and OPUS no.\ 2021/43/B/ST2/00497. M.~C., J. D. and A.~S. acknowledge financial support by the U.S. Department of Energy, Office of Nuclear Physics, Early Career Award under Grant No.\ DE-SC0020405.
J. D. also received support by the U.S. Department of Energy, Office of Science, Office of Nuclear Physics, within the framework of the TMD Topical Collaboration. 
The work of A.~M. has been supported by the National Science Foundation under grant number PHY-2110472, and also by the U.S. Department of Energy, Office of Science, Office of Nuclear Physics, within the framework of the TMD Topical Collaboration. 
F.~S.\ was funded by by the NSFC and the Deutsche Forschungsgemeinschaft (DFG, German Research Foundation) through the funds provided to the Sino-German Collaborative Research Center TRR110 “Symmetries and the Emergence of Structure in QCD” (NSFC Grant No. 12070131001, DFG Project-ID 196253076 - TRR 110). YZ was partially supported by an LDRD initiative at Argonne National Laboratory under Project~No.~2020-0020.
Computations for this work were carried out in part on facilities of the USQCD Collaboration, which are funded by the Office of Science of the U.S. Department of Energy. 
This research used resources of the Oak Ridge Leadership Computing Facility, which is a
DOE Office of Science User Facility supported under Contract DE-AC05-00OR22725.
This research was supported in part by PLGrid Infrastructure (Prometheus supercomputer at AGH Cyfronet in Cracow).
Computations were also partially performed at the Poznan Supercomputing and Networking Center (Eagle supercomputer), the Interdisciplinary Centre for Mathematical and Computational Modelling of the Warsaw University (Okeanos supercomputer), and at the Academic Computer Centre in Gda\'nsk (Tryton supercomputer). The gauge configurations have been generated by the Extended Twisted Mass Collaboration on the KNL (A2) Partition of Marconi at CINECA, through the Prace project Pra13\_3304 ``SIMPHYS".
Inversions were performed using the DD-$\alpha$AMG solver~\cite{Frommer:2013fsa} with twisted mass support~\cite{Alexandrou:2016izb}. 

\newlength{\bibitemsep}\setlength{\bibitemsep}{.2\baselineskip plus .05\baselineskip minus .05\baselineskip}
\newlength{\bibparskip}\setlength{\bibparskip}{0pt}
\let\oldthebibliography\thebibliography
\renewcommand\thebibliography[1]{%
  \oldthebibliography{#1}%
  \setlength{\parskip}{\bibitemsep}%
  \setlength{\itemsep}{\bibparskip}%
}

\end{document}